\documentclass[%
 reprint,
 amsmath,amssymb,
 aps,
pra,
]{revtex4-2}
\usepackage{booktabs}
\usepackage{graphicx}
\usepackage{dcolumn}
\usepackage{bm}
\usepackage{lipsum}
\usepackage{multirow}
\usepackage{xcolor}
\usepackage{relsize}
\usepackage{siunitx}
\usepackage{xparse}
\usepackage{array}
\usepackage{textgreek}

\begin{document}

\preprint{APS/123-QED}

\title{Electronic properties of the Radium-monochalcogenides RaX (X = O,S,Se) and RaO$^{\pm}$ ions}

\author{Mateo Londoño}
\author{Jesus Pérez-Ríos}
\affiliation{Department of Physics and Astronomy, Stony Brook University, Stony Brook, NY 11794, USA}


\begin{abstract}

We present a theoretical investigation on the electronic structure and properties of radium monochalcogenides, with chalcogens O, S, and Se, as well as the ionic species RaO$^{\pm}$. Our approach combines fully relativistic and partially relativistic quantum-chemistry methods. . Electronic properties are obtained using the exact two-component Hamiltonian–based coupled-cluster approach with single, double, and perturbative triple excitations [CCSD(T)+X2C], while potential energy curves are computed using an internally contracted multireference configuration interaction method, including relativistic effects through small-core pseudopotentials and Pauli--Breit operator diagonalization (MRCI+Q+ECP+SO). The dimers exhibit very large permanent dipole moments and sizable dipolar polarizabilities, while the Franck-Condon factors among the lowest electronic states are highly non-diagonal. These features are discussed in terms of the divalent character of the chemical bonding in the neutral species.

\end{abstract}

--\maketitle

\section{Introduction}
Diatomic molecules containing heavy radioactive atoms represent a unique scenario to probe new physics associated with parity (P) and time-reversal (T) symmetry–violating effects \cite{Safronova2017,Andreev2018,Kron2024}. In that respect, Radium-containing diatomic molecules, RaX, where X belongs to group 16 or 17, have received considerable attention \cite{Isaev2010,Isaev2020}. Experimentally, the developments of Garcia-Ruiz \textit{et al. } in the spectroscopy of short-lived radioactive molecules have opened the possibility of studying these molecules in the laboratory \cite{Ruiz2019}. Such spectroscopic studies greatly benefit from high-accuracy \textit{ab initio} electronic structure calculations. In particular, methods that describe accurately the electronic correlations and important relativistic effects of such heavy nuclei are required for guide experimental realizations \cite{Dyall2007,Liu2020, Alexander2020}. In this context, several studies have explored the electronic structure of radium monohalides, such as RaF \cite{Isaev2010, Kudashov2014, Zaitsevskii2022}, RaCl \cite{Isaev2020}, RaBr \cite{yuliya2021}, and RaI \cite{yuliya2022}, establishing their suitability for laser-cooling schemes \cite{Tarbutt2018}, among other key spectroscopic and structural properties. Furthermore, the triatomic RaOH molecule \cite{zhang2023, osika2024} has also been examined and identified as a promising candidate for laser cooling and for studies of symmetry-violating effects.

Regarding radium monochalcogenides, molecules formed by radium and group 16 elements, considerably fewer studies have been reported. Radium monochalcogenides belong to the broader class of alkaline-earth-metal monochalcogenides (AEM–MX), which have been extensively investigated for AEM elements ranging from Be to Ba, with MX = O and S \cite{Khalil2013,Ferreira2020,adamovic1999,KHATIB2017,KHATIB2018}. . From these studies, the general features of the ionic and divalent character of the chemical bonding are well established and consistent across the series. However, owing to the strong relativistic effects associated with Radium, the Ra–MX systems (MX = O, S, Se) have remained comparatively less explored. The existing literature is limited to studies focusing on the calculation of $P,T$-odd parameters in the ground state of RaO using single-reference relativistic coupled-cluster methods~\cite{Flambaum2007,Kudashov2013,Flambaum2020}.

In this work, we theoretically investigate the electronic structure and properties of RaX molecules with chalcogen atoms (X = O, S, Se) and the ions RaO$^{\pm}$. Our methodology incorporates several \textit{ab initio} quantum chemistry methods that allow us to account for partial electronic correlations and relativistic effects while maintaining computational feasibility. We compute the potential energy curves, dipole moments, static and dynamical dipolar polarizabilities, and the dispersion coefficients of the molecules. These results provide new insights into the chemical bonding of these exotic molecules and how it correlates with their large dipole moments and suitability for laser cooling.  Furthermore, the impact of relativistic effects on the various molecular properties is carefully analyzed.

The article is structured as follows: Sections \ref{sec:states} and \ref{sec:dipolar} focus on the description of the electronic states study and on the definition of dipole moments, polarizabilities, and dispersion coefficients, as computed here. In section \ref{sec:methods}, we described in detail the methods implemented for the calculation of the electronic potential energy curves and properties of the molecules, as well as a review of the basis sets implemented in the calculations. Section \ref{sec:results} is devoted to presenting the results and discussing them in the light of previous results in other AEM-MX studies, to emphasize the new insights gained with our study.






\section{Molecular orbitals and electronic states}
\label{sec:states}

In this section, we describe the electronic states calculated in this work. Throughout all the sections, the $\Lambda-S$ notation of the electronic terms is implemented. The coupled $\Omega$ notation will be used when discussing spin-orbit coupling in the coming sections.

The electropositive nature of alkaline earth metals (AEM) in combination with the high electronegativity of monochalcogenide atoms (MX) results in a polar dominant character of the bonding in heteronuclear molecules of these two species. In particular, the dominant binding mechanism  in such a system involves the charge transfer of the two external electrons of the AEM to the MX. This is widely supported by previous calculations of the electronic states of AEM+MX molecules\cite{Khalil2013,Ferreira2020,adamovic1999}. This special bonding mechanism, often implies that the diatomic electronic ground state does not correlate to the asymptotic atomic ground state of the AEM+MX system. Based on these observations, we decided to explore the electronic states associated with the three lowest atomic asymptotes of the Ra-X (X=O,S,Se) system, including the molecular ground states, the atomic ground state asymptote and an additional set of excited states.

Table~\ref{tab:states_RaO} displays the atomic asymptote states and the resulting $\Lambda$-S molecular states for RaO. Both excited asymptotes come from the excited states of Radium: Ra($^{3}$P$^{\circ}$) and Ra($^{3}$D) states, which are 13078.40 cm$^{-1}$ and 13715.779 cm$^{-1}$ above the ground state asymptote. Altogether, these limits result in a total of 44 $\Lambda$-S states, between singlets, triplets, and quintuplets. The next excited asymptote correlates with the ground Ra atom and the O($^{1}$D), which is 15867.862 cm$^{-1}$ above the ground states and is not included here. The first ionization energies for Ra and O are 5.2784 eV and 13.6181 eV, respectively, allowing us to neglect ionic channels in this first calculation.

\begin{table}[!ht]
\centering
\begin{tabular}{l>{\centering\arraybackslash}m{4.9cm}>{\centering\arraybackslash}m{1.4 cm}}
\hline
\textbf{Channels} & \textbf{Electronic States} & $\mathbf{C_{2v}}$ \textbf{Irrep.} \\
\hline
\hline
\\
 Ra($^1$S)+O($^3$P) & {$^3\Sigma^-$, $^3\Pi$} & {(0,1,1,1)} \\
\hline
\\
Ra($^3$P$^{\circ}$)+O($^3$P) &{$^{1,3,5}\Sigma^+$, $^{1,3,5}\Sigma^-$(2), $^{1,3,5}\Pi$(2), $^{1,3,5}\Delta$} &{(2,2,2,3)} \\
\hline
\\
Ra($^3$D)+O($^3$P) & $^{1,3,5}\Sigma^+$, $^{1,3,5}\Sigma^-$(2), 2$^{1,3,5}\Pi$(3),  & {(3,4,4,4)} \\
 & $^{1,3,5}\Delta$(2), $^{1,3,5}\Phi$ & \\
\hline
\end{tabular}
\caption{Molecular channels and corresponding $\Lambda$–$S$ electronic states for RaO, and its $C_{2v}$ irreducible representations.}
\label{tab:states_RaO}
\end{table}

The atomic asymptotes and molecular states for RaS and RaSe are shown in Table~\ref{tab:states_RaS_RaSe}. Both molecules present equivalent asymptotic channels and, hence, the same set of electronic states, which in this case corresponds to a total of 23 $\Lambda$-S states, including all spin multiplicities. The excited asymptotic channels arise from the MX ($^{1}$D) states, which are 9238.609 cm$^{-1}$ and 9576.149 cm$^{-1}$ above the ground-state threshold for S and Se, respectively, and from the first excited state of Ra, Ra($^{3}$P$^{\circ}$). In this case, the fourth asymptotic channel, due to the second excited state of Ra, Ra($^{3}$D), is not considered. Also, as before, we point out that the first ionization potentials of S and Se are 10.360 eV and 9.7524 eV, respectively, which are large enough to disregard these channels.

Here, we present the results only for 10 of the $\Lambda$-S states of each molecule. The states included the $\Sigma^{\pm}, \Delta, \Phi$ symmetry states for singlets and triplets, while quintuplet states are all dissociative and higher in energy, and we do not analyze them in this study. For the spin-orbit states, attention is paid only to the lowest energy states, and the grid of internuclear distance is concentrated in the region around the minima, without resolving the asymptotic behavior of the PES.




\begin{table}[!ht]
\centering
\begin{tabular}{l>{\centering\arraybackslash}m{4.9cm}>{\centering\arraybackslash}m{1.4 cm}}
\hline
\textbf{Channels} & \textbf{Electronic States} & $\mathbf{C_{2v}}$ \textbf{Irrep.} \\
\hline
\hline

Ra($^1$S)+S($^3$P) & \multirow{2}{*}{$^3\Sigma^-$, $^3\Pi$} & \multirow{2}{*}{(0,1,1,1)} \\
Ra($^1$S)+Se($^3$P) & & \\
\hline
Ra($^1$S)+S($^1$D) & \multirow{2}{*}{$^1\Sigma^+$, $^1\Pi$, $^1\Delta$} & \multirow{2}{*}{(2,1,1,1)} \\
Ra($^1$S)+Se($^1$D) & & \\
\hline
Ra($^3$P$^{\circ}$)+S($^3$P) & \multirow{2}{*}{$^{1,3,5}\Sigma^+$, $^{1,3,5}\Sigma^-$(2), $^{1,3,5}\Pi$(2), $^{1,3,5}\Delta$} & \multirow{2}{*}{(2,2,2,3)} \\
Ra($^3$P$^{\circ}$)+Se($^3$P) & & \\
\hline
\end{tabular}
\caption{Molecular channels and corresponding $\Lambda$–$S$ electronic states for RaS and RaSe, and its $C_{2v}$ irreducible representations.}
\label{tab:states_RaS_RaSe}
\end{table}

Moreover, we consider the anion and cation states of the RaO molecule. The ground state of the cation RaO$^{+}$, correlates with the asymptotic channel Ra$^{+}$($^{2}$S) + O($^3$P). In this case the bonding is mainly ionic in character but only one electron of the Ra$^{+}$ is transferred to O. We analyze the first two excited asymptotic states. The first one arising from the $7s \rightarrow 6d$ excitation of the valence electron in Ra$^{+}$, which is at 12084 cm$^{-1}$ above the ground asymptote, and the second excited state from the O($^{1}$D) configuration of Oxygen at 15867 cm$^{-1}$ above the fundamental threshold. These atomic terms along with the molecular states are tabulated in table \ref{tab:states_RaO_ions}.

On the other hand, for the anion RaO$^{-}$ we consider only the ground state threshold. The electron affinity of oxygen is larger than that of Ra, and the ground atomic channel is Ra($^{1}$S)+O($^{2}$P$^{\circ}$). The resulting electronic states are compiled in table \ref{tab:states_RaO_ions}. We note that the dipole moment of RaO is larger than 2~D and dipole bound states of the RaO$^{-}$ ion may exist, but they are not addressed in this work.

\begin{table}[ht!]
\centering
\begin{tabular}{l>{\centering\arraybackslash}m{4.9cm}>{\centering\arraybackslash}m{1.4 cm}}
\hline
\textbf{Channels} & \textbf{Electronic States} & $\mathbf{C_{2v}}$ \textbf{Irrep.} \\
\hline
\hline
\\
&\textbf{RaO}$^{+}$&\\
\hline
\hline
\\
 Ra$^{+}$($^2$S)+O($^3$P) & {$^{2,4}\Sigma^-$, $^{2,4}\Pi$} & {(0,1,1,1)} \\
\hline
\\
Ra$^{+}$($^2$D)+O($^3$P) &$^{2,4}\Sigma^+$, $^{2,4}\Sigma^-$(2), $^{2,4}\Pi$(3),  &{(3,4,4,4)} \\
& $^{2,4}\Delta$(2),$^{2,4} \Phi$ &\\
\hline
\\
Ra$^{+}$($^2$S)+O($^1$D) & $^{2}\Sigma^+$, $^{2}\Pi$, $^{2}\Delta$ & {(3,4,4,4)} \\
\hline
\hline
\\
&\textbf{RaO}$^{-}$&\\
\hline
\hline
\\
Ra($^1$S)+O$^{-}$($^2$P$^{\circ}$) & $^{2}\Sigma^+$, $^{2}\Pi$  & {(1,1,1,0)} \\
\hline
\end{tabular}
\caption{Molecular channels and corresponding $\Lambda$–$S$ electronic states for RaO$^{\pm}$ ions, and its $C_{2v}$ irreducible representations.}
\label{tab:states_RaO_ions}
\end{table}

\section{Polarizabilities and dipole moments}
\label{sec:dipolar}
The static polarizability and dipole moment are fundamental quantities to describe the interaction of the RaX system with an external static field \textbf{E}. In the presence of the component $E_{j}$, of the field in the direction $j$, the electronic energy $U$ of the molecule is modified and can be expanded as
\begin{equation}
    U(E_{j}) = U_{0} + \frac{dU}{dE_{j}}\Biggr|_{\substack{E_j=0}} E_{j}  + \frac{1}{2}\frac{d^{2}U}{dE_{j}^{2}}\Biggr|_{\substack{E_j=0}}E_{j}^{2} + ...,
\end{equation}
where $U_{0}$ is the non-perturbated electronic energy of the molecule and the coefficients of the linear and quadratic terms define the $j$-component of the dipole moment and the dipole polarizability as
\begin{equation}
    d_j = \frac{dU}{dE_{j}}\Biggr|_{\substack{E_j=0}}  \,\,\,\, , \,\,\,\,  \alpha_{jj}=\frac{d^{2}U}{dE_{j}^{2}}\Biggr|_{\substack{E_j=0}},
    \label{eq:eq2}
\end{equation}
respectively. For heteronuclear linear diatomic systems oriented along the $z$ axis, the relevant dipole moment component is $d_z$, and we define the parallel ($\alpha_\parallel$) and perpendicular ($\alpha_\perp$) components of the polarizability tensor as
\begin{equation}
    \alpha_\parallel = \alpha_{zz} \,\,\,,\,\,\, \alpha_\perp=\frac{1}{2}(\alpha_{xx} + \alpha_{yy}).
\end{equation}
The mean polarizability of the system is defined as $\bar{\alpha} = (\alpha_\parallel + 2\alpha_\perp)/3 $.

In addition to the static properties, we evaluated the dynamic polarizability at imaginary frequencies, $\alpha(i\omega)$ for the neutral systems RaX, which allows us to compute the dispersion coefficient $C_{6}^{disp}$ for the interaction of the dimers by means of the Casimir-Polder integral given by
\begin{equation}
    C_{6}^{disp} =  \frac{3}{\pi}\int_{0}^{\infty}\bar{\alpha}_{\rm{RaX}}(i\omega)\bar{\alpha}_{\rm{RaY}}(i\omega)d\omega.
    \label{eq:eq4}
\end{equation}
In this work we evaluate the electronic properties and dispersion coefficients for the electronic ground state of the systems under consideration, X$^{1}\Sigma^{+}$.

\section{Computational details}
\label{sec:methods}
The Ra–X systems considered in this work are strongly relativistic molecules. For related systems, a variety of approaches based on fully relativistic four-component Hamiltonians or accurate exact two-component (X2C) formulations have been employed to compute electrostatic properties and potential energy surfaces \cite{Isaev2010, Kudashov2014, Zaitsevskii2022,Isaev2020,yuliya2021,yuliya2022}. In particular, the Fock-space relativistic coupled-cluster (FS-RCC) method has been successfully applied to describe the electronic structure of RaF, RaOH, and RaCl, among others. However, the fully relativistic all-electron treatment inherent to these approaches renders them computationally very demanding \cite{FLEIG2012,CHENG2024}. As an alternative to fully relativistic methodologies, the use of small-core relativistic pseudopotentials offers a practical route to reduce computational cost, albeit at the expense of some loss of accuracy \cite{stoll2002}.

Here, the permanent dipole moments and dipole polarizabilities are computed using a full-electron X2C Hamiltonian in combination with suitable basis sets and a finite-field approach to evaluate the derivatives appearing in Eq. \ref{eq:eq2}. For the potential energy surfaces, instead of employing the fully relativistic FS-RCC methodology, we adopt a computationally more affordable alternative based in a multireference configuration interation approach that incorporates relativistic effects via small-core, energy-consistent pseudopotentials and a perturbative treatment of spin–orbit coupling (MCSCF/MRCI+Q+ECP+SO). The accuracy and limitations of the resulting values are discussed with more detail in section \ref{sec:appendix}.

\subsubsection{$\Lambda$-S PES }

For the systems under consideration, we use the ECP78MDF small-core pseudopotential to describe the 78 core electrons of Ra, including relativistic effects, and the aug-cc-pVQZ-PP basis set with $spdfg$ functions to model the remaining 10 electrons. The accuracy of this approach has been tested against the spectroscopic constants of RaF, and the results are shown in the appendix (Section~\ref{sec:appendix}). In the appendix, we also provide further calculations on molecules such as BaF and BaO and compare the results with more sophisticated numerical results and experimental data.

Given that MOLPRO does not support the infinite point group $C_{\infty v}$, the calculations were carried out in the $C_{2v}$ point group. The irreducible representations in this group consist of four symmetries: $A_{1}$, $B_{1}$, $B_{2}$, and $A_{2}$. The molecular orbitals (MOs) are classified as: $\sigma \equiv A_{1}$, $\pi,\phi \equiv B_{1} + B_{2}$, and $\delta \equiv A_{1} + A_{2}$. Similarly, the electronic states correspond to: $\Sigma^{+} \equiv A_{1}$, $\Pi,\Phi \equiv B_{1} + B_{2}$, $\Delta \equiv A_{1} + A_{2}$, and $\Sigma^{-} \equiv A_{2}$.

\subsection{RaO}
For oxygen (O), we used Dunning’s augmented correlation-consistent, polarization-weighted basis set with five-zeta functions: aug-cc-pwCV5Z, to describe $s,p,d,f$ and $g$ functions. In the calculations, all of the 8 electrons in the electronic configuration $1s^22s^22p^4$ of oxygen were included, together with the 10 electrons of Ra that are not included in the effective core potential (ECP78MDF). As a result, the total ground-state orbital space is described by the irreducible representation (6,2,2,0). The inclusion of the two excited atomic asymptotes of the system, add the orbitals $7p$ and $6d$ of radium. Furthermore, the next excited asymptote correlates with the first exited state of oxygen O($^{1}$D), which is still describe by the (2$s^{2}$2$p^{4}$) configuration; thus, no further orbitals are added, and the total orbital space is given by (9,4,4,1). From this set of orbitals, the $1s$ orbital of O and the $6s6p$ orbitals of Ra, correspond to the close-orbital space (3,1,1,0), while the rest of the orbitals belong to the active space(6,3,3,1). Initial MCSCF calculations, varying the active space verify the contribution from the above mentioned orbitals and then confirm our initial choice.

\subsection{RaS}

In the case of sulfur (S), we use the aug-cc-pwCV5Z basis set to describe $spdf$ functions. All 16 electrons are described and included in the basis set. The valence orbital space incorporates the $3s3p$ electrons of S and the $7s$ electrons of Ra. Additionally, the excited atomic asymptotes result in the inclusion of the $7p$ orbitals for Ra. Initial MCSCF calculations show that the $6d$ orbitals of Ra also display some contributions to the lower electronic states, as expected due to the small difference in energy between the $^3$P$^{\circ}$ and the $^3$D of Ra (637.37 cm$^{-1}$) and the high ionic character of the bond. Then the active space has 8 electrons in a total of 12 orbitals with irreducible representation (6,3,3,1). The rest of electrons in the calculations belong to the core space, (5,2,2,0).



\subsection{RaSe}
Selenium (Se) has a total of 34 electrons. To make the calculations for RaSe computationally affordable while including atomic relativistic effects, we modeled 10 core electrons of selenium using the ECP10MDF small-core relativistic and energy-consistent pseudopotential. The remaining 24 electrons, corresponding to the electronic configuration $3s^23p^63d^{10}4s^24p^4$, were included in the orbital space and described using the augmented correlation-consistent polarized valence basis set with five-zeta functions, aug-cc-pV5Z-PP, optimized for the chosen pseudopotential. Along with the 10 electrons of radium and the excited asymptotic states, the total orbital space is described by the irreducible representation (12,6,6,2). The core orbitals included the $3 s,p,d$ orbitals of Se and the usual $6s,p$ of Ra.

\subsection{RaO$^{\pm}$}
For the ionic states of radium monoxide, we still used the same basis sets and pseudopotentials. The active space, however, changes, leading to a more compact one compared to the neutral species. In the case of RaO$^{+}$ we include all the electrons in the $6s^26p^67s^1$ configuration of Ra$^{+}$ and additionally the $6d$ orbitals to describe the first excited state Ra$^{+}(^2$D). This choice, along with the electronic configuration of oxygen, results in a total orbital space of (8,3,3,1), and closed orbitals $6sp$ of Ra and $1s$ of O, (3,1,1,0). On the other hand, for RaO$^{-}$, we consider the 10 pseudo-core and valence electrons of Ra($^{1}$S) and the 9 electrons of the anion O$^-$($^2$P$^\circ$). Including only the ground channel, we have the total orbital space (6,2,2,0), and close again the $6sp$ of Ra and $1s$ of O.

\subsection{CCSD(T) and CCSD-EOM}

In addition to the uncertainties arising from the non-fully relativistic treatment inherent to the MRCI+Q+ECP+SO approach employed in this work, there is a further source of error associated with the description of dynamical electron correlation within the MRCI+Q framework. To assess the accuracy of the reported results, we performed additional benchmark calculations using coupled-cluster theory with single, double, and perturbative triple excitations, CCSD(T). These calculations were carried out using the same basis sets and small-core energy-consistent pseudopotentials described above, and were used to determine equilibrium bond lengths and excitation energies for the $X^{1}\Sigma^+ \rightarrow a^3\Pi$ transitions.

Furthermore, excitation energies for the singlet–singlet transitions $X^{1}\Sigma^+ \rightarrow A^1\Pi$ were computed using the equation-of-motion coupled-cluster method with single and double excitations (EOM-CCSD). All coupled-cluster calculations were performed with the MOLPRO package. For the EOM-CCSD results, we expect an accuracy comparable to that of CCSD for excited states predominantly characterized by single excitations, as is the case for the transitions considered here.

\subsubsection{Spin-Orbit calculations}

Spin-orbit coupling is included in the calculations by using the one- and two-electron Pauli–Breit (PB) operators at the MRCI+Q level of theory, with the same basis sets and pseudopotentials as before. In this calculation, a reduced number of the lower $\Lambda-S$ states is used to diagonalize the PB operator on an internuclear grid around the equilibrium distance of the ground electronic state. For the new $\Omega$-states we report the potential curves and spectroscopic constant values.

\subsubsection{Static polarizabilities and dipole moments}
The values for the dipole moment and static dipole polarizability components were computed as given by equation \ref{eq:eq2}, and the derivatives were computed by implementing a finite-field (FF) approach, which evaluates the derivatives in a finite-difference scheme as
\begin{equation}
    d_j = \frac{U(E_{j})-U(-E_{j})}{2E_{j}}\,,\,  \alpha_{jj}=\frac{U(E_{j})+U(-E_{j}) -2U(0)}{E_{j}^{2}}.
    \label{eq:eq5}
\end{equation}
Two different approaches computed the electronic energies of the system. First, the CCSD(T)+ECP method, where the electronic energies are computed at the CCSD(T) level of theory, and relativistic effects for Ra are, once again, included with fully relativistic small-core pseudopotential ECP78MDF. The second approach implements an exact two-component (X2C) Hamiltonian and determines the electronic energy using the CCSD(T) approach. These calculations were carried out in the DIRAC 23.0 software \cite{DIRAC23} as well as in MOLPRO. For each software a different basis set was implemented.



In the MOLPRO implementation of the CCSD(T)+X2C method, Ra was modeled with the all-electron augmented correlation-consistent quadruple-zeta basis set aug-cc-pVQZ-X2C, specifically designed for use with exact two-component relativistic Hamiltonians. The basis was used in its spherical harmonic representation and in its contracted form
(38s,35p,25d,18f,2g)/[11s,10p,8d,5f,2g], and $spdfg$ functions were considered. For O and S, we used $spdfg$ functions of the aug-cc-pVQZ basis set. For Se, the relativistic all-electron augmented correlation-consistent quadruple-zeta basis set aug-cc-pVQZ-X2C is implemented in the contracted form (22s,17p,13d,3f,2g)/[8s,7p,5d,3f,2g]. In the case of the DIRAC software implementation, we have use the all electrons relativistic Dyall’s relativistic augmented valence n-zeta basis sets, dyall.avnz, with n=2,3; for all the atoms involved in the calculations. In this case, we explicitly include only valence-valence and sub-valence-valence electronic correlation.

\subsubsection{Dispersion coefficients}

The dynamic polarizabilities at imaginary frequencies were computed for the neutral species using the time-independent coupled-cluster polarization propagator in the singles and doubles approximation (TI-CCSD), as implemented in MOLPRO. Relativistic effects are included via an energy-dependent pseudopotential. For the TI-CCSD+ECP method, the basis set choices are similar to those of the CCSD(T)+ECP method for static polarizations.

Once the values of dynamic polarizability at imaginary frequencies are extracted, the dispersion coefficients are computed by numerical evaluation of the Casimir-Polder integral \ref{eq:eq4}, using a 50-points Gauss-Legendre as suggested in reference \cite{DEREVIANKO2010} .

\section{Results}
\label{sec:results}
\subsection{Spin-free calculations}

\subsubsection{Ra-X}
\begin{figure*}[htbp!]
\includegraphics[width=\textwidth]{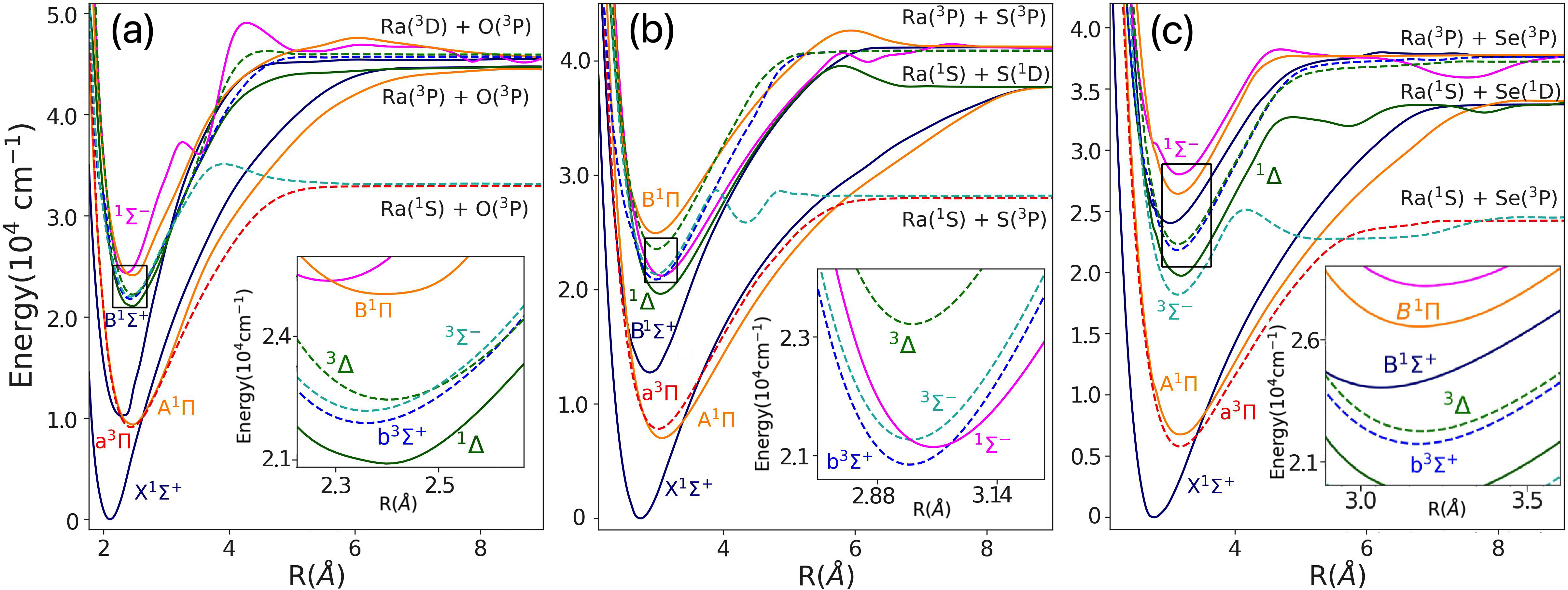}  
\caption{ The MRCI+Q + ECP potential energy curves of 10  lowest $\Lambda$-S electronic states for the neutral Radium-monochalcogenides: (a) RaO, (b) RaS and (c) RaSe . Solid lines are used for the singlet states, while dashed lines refers to triplet states. The inset zoomed the region in the black box.}
\label{fig:PES_RAX}
\end{figure*}

The $\Lambda-S$ potential energy curves for some of the electronic states for RaO, RaS, and RaSe are shown in panels (a), (b), and (c) of Fig.~\ref{fig:PES_RAX}, respectively. It is easy to see the resemblance among the three molecules, in which the major differences reside in the energy difference between the atomic channels. However, despite the similarity in the shape of the electronic energy curves, it is necessary to perform a case-by-case study.

\begin{figure}[h]
\centering
 \includegraphics[width=\linewidth]{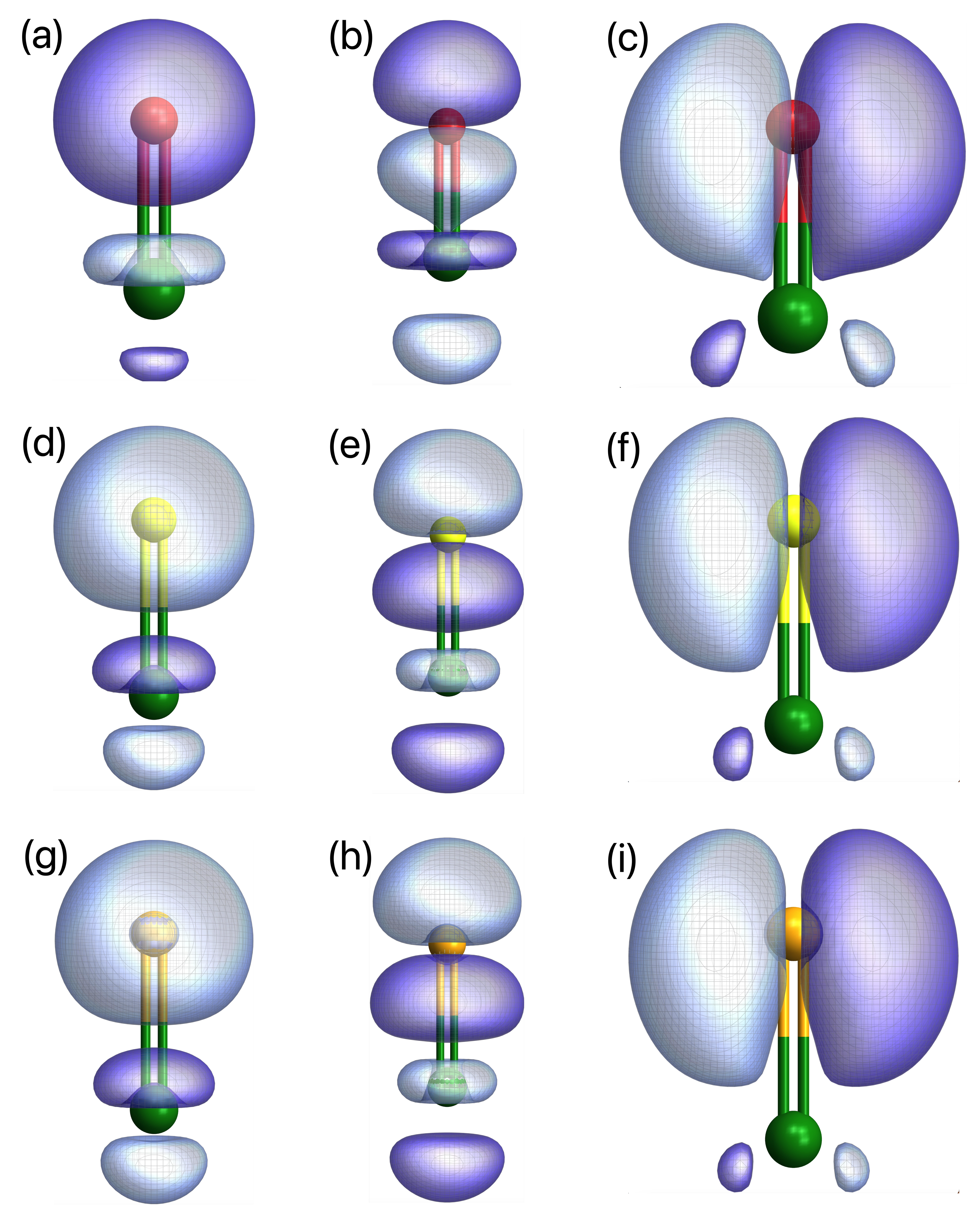}  
\caption{Molecular orbitals for the valence electron in RaO (a,b,c), RaS (d,e,f) and RaSe (g,h,i). The orbitals represents in order the 1$\sigma$,2$\sigma$ and 3$\pi_{1}$ orbitals constituting the X$^{1}\Sigma^{+}$. In all the cases the green atom is Ra, and the divalent stucture of the bond is represented. }
\label{fig:orbitals}
\end{figure}

In the case of RaO, the figure shows the dominant electronic configurations of the four lowest states, as listed in Table~\ref{tab:configurations}. Here, we only enumerate the molecular orbitals of the 8 valence electrons in the system, so that the ground state has the configuration $1\sigma^{2}2\sigma^{2}1\pi^{4}$. The bivalent ionic binding of the molecule leads to an $X^{1}\Sigma^{+}$ ground state, related to the first excited atomic asymptote Ra($^3$P$^{\circ}$)+O($^3$P). The molecular orbitals in the $1\sigma^{2}2\sigma^{2}1\pi^{4}$ configuration of the ground state have their main contributions from the $2s$ and $2p$ atomic orbitals of oxygen. The $2s$, $2p_{z}$, and $2p_{x,y}$ orbitals of oxygen dominate the $1\sigma$, $2\sigma$, and $1\pi$ molecular orbitals, resulting in a significant transfer of the two $7s$ electrons of Ra to the two $2p$ holes in O. This is better highlighted in panels (a) to (c) of Fig.~\ref{fig:orbitals}, showing that the orbitals are mainly centered on O. For instance, the molecular orbital $2\sigma$ in panel (b) resembles a $p_{z}$ atomic orbital of oxygen. The first excited triplet and singlet states, $a^{3}\Pi$ and $A^{1}\Pi$, are located at around 10000 cm$^{-1}$ above the ground state, sharing similar electronic configurations around the equilibrium distance, as shown in Table~\ref{tab:configurations}. The excited configuration involves the $3\sigma$ molecular orbital, which is dominated by the $7s$ atomic orbital of Ra. Given this common electronic configuration between both states, strong SO mixing arises and mixes both states around 2.8~\AA. The next singlet excited state, $B^{1}\Sigma^{+}$, is 10234.01 cm$^{-1}$ above the first excited state. The spectroscopic constants for the lowest RaO molecular states are tabulated in Table~\ref{tab:spectroscopy}, along with different values for comparison.


We estimate the error in the equilibrium bond length relative to CCSD(T)+ECP to be approximately 1.15\%. The calculated ground-state bond length is about 0.06~\AA\ larger than the value reported in Ref.~\cite{Kudashov2013}. For the energy separation between the first singlet and triplet states, the deviation from the coupled-cluster result is 8.12\%. In the case of the singlet–singlet transition, the discrepancy with respect to CCSD-EOM+ECP is 17.65$\%$. It is worth noting that, as discussed in Section~\ref{sec:appendix}, and consistent with previous experience for BaO, the CCSD-EOM+ECP method for AEM+O systems tends to overestimate the lowest singlet–singlet excitation energy.



The electronic ground state for RaS, $X^{1}\Sigma^{+}$, is dominated by the electronic configuration $1\sigma^{2}2\sigma^{2}1\pi^{4}$, as for RaO. However, in this case, the valence molecular orbitals are not entirely dominated by the chalcogen contribution. In particular, the $1\sigma$ orbital has a significant contribution from the Ra $6p_{z}$ atomic orbital. Because sulfur has a lower electron affinity, the radium atom can attract electron density more strongly, leading to noticeable differences in the molecular orbitals. These differences are illustrated in figure \ref{fig:orbitals} (panels d-f). This behavior also results in an additional 22.20$\%$ contribution from the configuration $1\sigma^{2}2\sigma^{\alpha}1\pi^{4}3\sigma^\beta$ to the molecular ground state, where the $3\sigma$ is dominated by the $7s$ orbital of Ra. The first excited triplet state, a$^3 \Pi$ is at 7846 cm$^{-1}$ above the ground state, while A$^1\Pi$ is lower in energy, at approximately 6000 cm$^{-1}$. The B$^{1}\Sigma^+$ is now energetically seprated from the lowest excited states, appearing at 12753.58 cm$^{-1}$ above the ground state.



\begin{table}[h!]
\centering
\begin{tabular}{l>{\centering\arraybackslash}m{4.0 cm}>{\centering\arraybackslash}m{2.0 cm}}
\hline
\textbf{State} & \textbf{MRCI-configuration} & \textbf{Coefficient} \\
\hline
\hline
 & \textbf{         RaO} &
\\
\hline
\hline
\\
 X$^{1}\Sigma^{+}$ & ...1$\sigma^{2}2\sigma^{2}1\pi^{4}$    & 87.0$\%$ \\
 \hline
\\
 a$^{3}\Pi$ & ...1$\sigma^{2}2\sigma^{2}1\pi^{3(\alpha)}3\sigma^{\alpha}$& 96.4 $\%$\\
\hline
\\
 A$^{1}\Pi$ & ...1$\sigma^{2}2\sigma^{2}1\pi^{3(\beta)}3\sigma^{\alpha}$   & 96.4$\%$  \\
\hline
\\
 B$^{1}\Sigma^{+}$ & ...1$\sigma^{2}2\sigma^{\alpha}1\pi^{4}3\sigma^{\beta}$   & 95.0$\%$ \\
\hline
\hline
 & \textbf{         RaS} &
\\
\hline
\hline
\\
 X$^{1}\Sigma^{+}$ & ...1$\sigma^{2}2\sigma^{2}1\pi^{4}$,  ...1$\sigma^{2}2\sigma^{\alpha}1\pi^{4}3\sigma^{\beta}$  &  77.27$\%$, 22.20$\%$  \\
\hline
\\
 a$^{3}\Pi$ & ...1$\sigma^{2}2\sigma^{2}1\pi^{3(\alpha)}3\sigma^{\alpha}$& 80.83$\%$ \\
\hline
\\
 A$^{1}\Pi$ & ...1$\sigma^{2}2\sigma^{2}1\pi^{3(\beta)}3\sigma^{\alpha}$   &94.31 $\%$  \\
\hline
\\
 B$^{1}\Sigma^{+}$ & ...1$\sigma^{2}2\sigma^{\alpha}1\pi^{4}3\sigma^{\beta}$   & 84.44$\%$  \\
\hline
\hline
 & \textbf{         RaSe} &
\\
\hline
\hline
\\
 X$^{1}\Sigma^{+}$ & ...1$\sigma^{2}2\sigma^{2}1\pi^{4}$,  ...1$\sigma^{2}2\sigma^{\alpha}1\pi^{4}4\sigma^{\beta}$  &  77.89$\%$, 21.15$\%$  \\
\hline
\\
 a$^{3}\Pi$ & ...1$\sigma^{2}2\sigma^{2}1\pi^{3(\alpha)}3\sigma^{1}$& 96.34$\%$\\
\hline
\\
 A$^{1}\Pi$ & ...1$\sigma^{2}2\sigma^{2}1\pi^{3(\beta)}3\sigma^{\alpha}$   & 96.45$\%$ \\
\hline
\\
 B$^{1}\Sigma^{+}$ & ...1$\sigma^{2}2\sigma^{\alpha}1\pi^{3(\alpha)}2\pi_{1}^{\beta}$,
 ...1$\sigma^{2}2\sigma^{\alpha}1\pi^{3(\alpha)}2\pi_{-1}^{\beta}$
 &48.25$\%$ , 48.25$\%$ \\
\hline
\end{tabular}
\caption{Electronic configurations for the 4 lowest $\Lambda - S$ electronic states of the neutral Ra-X}
\label{tab:configurations}
\end{table}

For RaSe, the electronic ground state, $X^{1}\Sigma^{+}$, exhibits contributions from the same electronic configurations as RaS. In this case, the $1\sigma$ molecular orbital has its largest contribution from the Ra $6p_{z}$ atomic orbital. The orbitals shown in panels (g) to (i) of Fig.~\ref{fig:orbitals}, are very similar to the RaS, since the bonding mechanism is essentially the same for both molecules. The first two excited states lie closer to the ground state than in the previous molecules, with the $a^{3}\Pi$ state at $5979.822~\mathrm{cm}^{-1}$ and the $A^{1}\Pi$ state at $6764.435~\mathrm{cm}^{-1}$ above the $X^{1}\Sigma^{+}$ ground state. In contrast, the $B^{1}\Sigma^{+}$ state is much higher in energy, lying at $24\,039.00~\mathrm{cm}^{-1}$ above the ground state. This behavior arises because the $X^{1}\Sigma^{+} \rightarrow B^{1}\Sigma^{+}$ transition is now dominated by the $2\sigma \rightarrow 2\pi_{\pm}$ electronic excitation, which is more energetic than the $2\sigma \rightarrow 3\sigma$ transition found for RaO and RaS (see Table~\ref{tab:configurations}).

\begin{figure}[h]
\centering
\includegraphics[width=\linewidth]{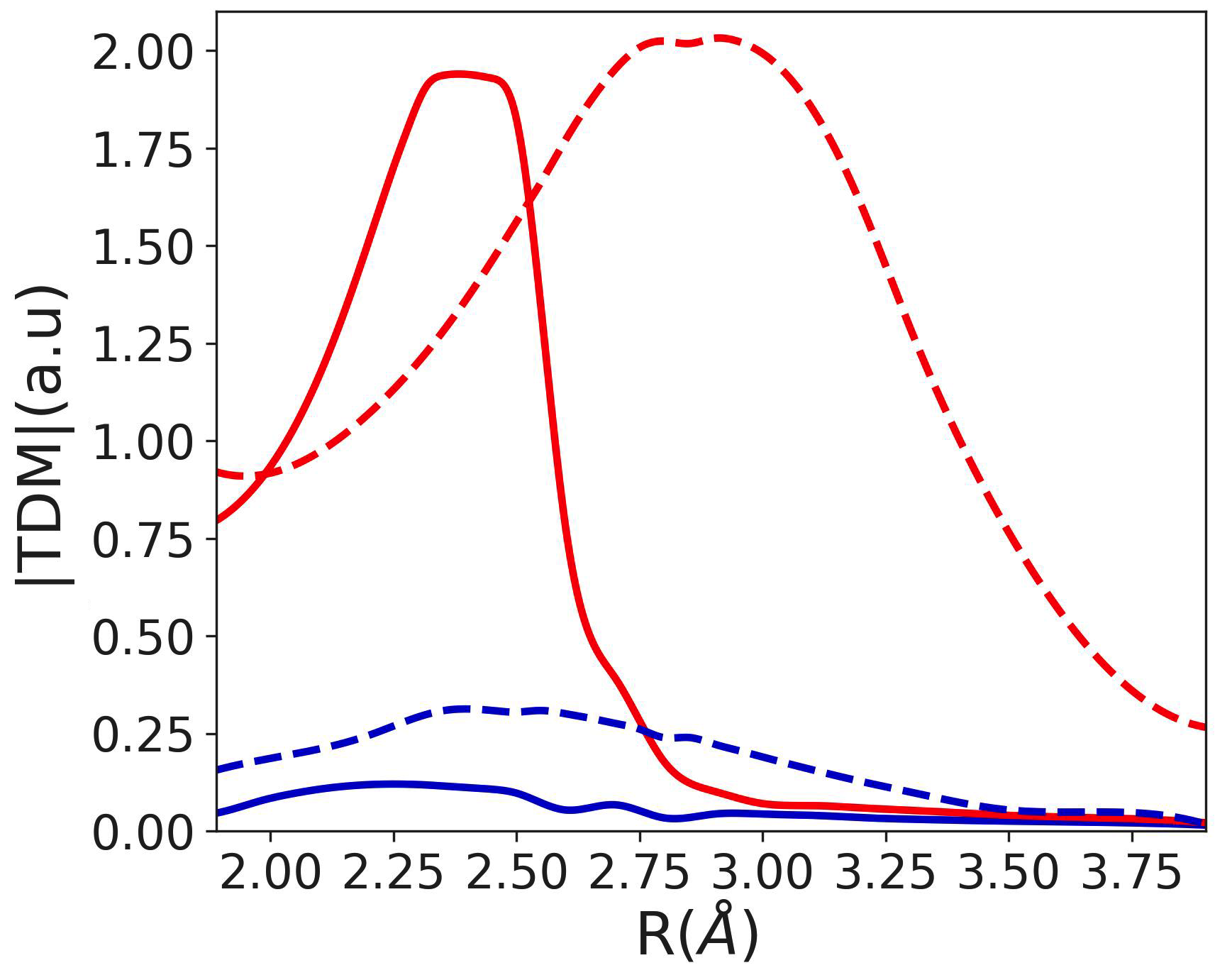}  
\caption{Absolute value of the transition dipole moment for the X$^{1}\Sigma^{+}\longrightarrow $B$^1\Sigma^+$ (blue curves) ,and X$^{1}\Sigma^{+}\longrightarrow $A$^1\Pi$ (blue curves) transitions, in the RaO (Solid lines) and RaS (dahsed lines) molecules.}
\label{fig:TDM}
\end{figure}

The absolute value of the transition dipole moment (TDM) for the X$^{1}\Sigma^{+}\longrightarrow $B$^1\Sigma^+$, and X$^{1}\Sigma^{+}\longrightarrow $A$^1\Pi$ transitions for RaO and RaS, in the region around the potential minima, is displayed in figure \ref{fig:TDM}. The TDMs for the singlet $\Sigma \longrightarrow \Sigma$ transition are larger than the $\Sigma \longrightarrow \Pi$ transitions, and around the equilibrium position of both $\Sigma$ potentials have a considerable value. In Fig.~\ref{fig:TDM} we only plot RaO and RaS since the results for RaSe are very similar to RaS.

\begin{figure*}[htbp!] %
  \includegraphics[width=\textwidth]{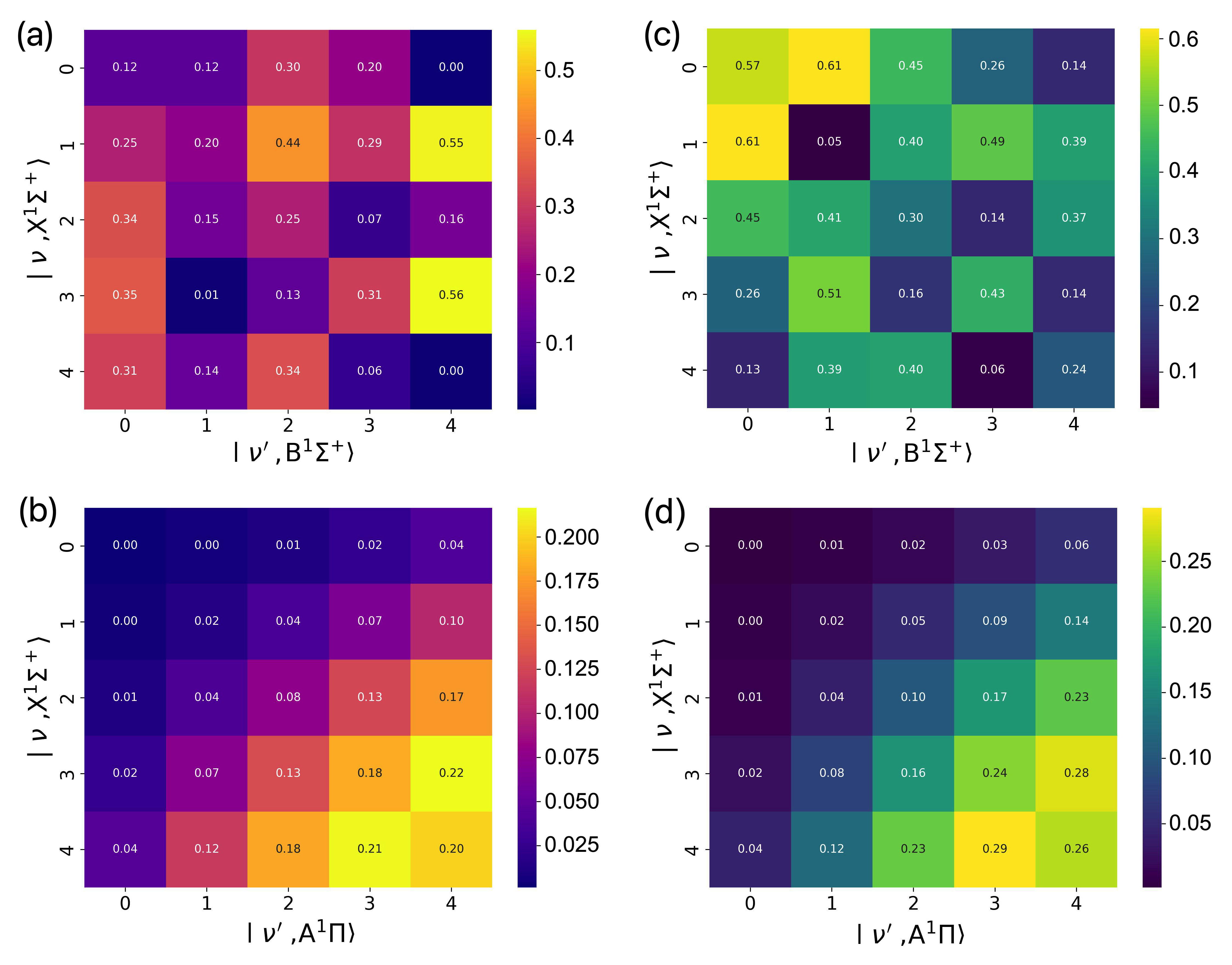} 
    \caption{Franck-Condon factors of the first five vibrational states for the single-single transitions: (a) X$^{1}\Sigma^{+}\longrightarrow $B$^{1}\Sigma^{+}$ of RaO, (b)X$^{1}\Sigma^{+}\longrightarrow $A$^{1}\Pi$ of RaO, (c) X$^{1}\Sigma^{+}\longrightarrow $B$^{1}\Sigma^{+}$ of RaS and (d)X$^{1}\Sigma^{+}\longrightarrow $A$^{1}\Pi$ of RaO }
    \label{fig:FCF}
\end{figure*}

We compute the vibrational energies and wavefunctions using a discrete-variable representation (DVR) method for the $X^1\Sigma^+$, $A^1\Pi$, and $B^1\Sigma^+$ electronic states of RaO and RaS. Using the resulting wavefunctions, we evaluate the Franck–Condon factors (FCFs) between the first five vibrational levels of the $X^1\Sigma^+ \rightarrow A^1\Pi$ and $X^1\Sigma^+ \rightarrow B^1\Sigma^+$ electronic transitions. The resulting FCFs are presented in matrix form in Fig.~\ref{fig:FCF}. First, we note that the Franck–Condon factors are relatively small, with maximum values of 0.61 for RaS and 0.56 for RaO. Second, the FCF matrices exhibit a clear absence of a diagonal structure. This behavior can be rationalized in light of the values of the spectroscopic constants displayed in Table~\ref{tab:spectroscopy}, which directly affect the overlap of the vibrational wavefunctions. These findings strongly contrast with those observed in Ra monohalide molecules such as RaF and RaCl, where highly diagonal FCF matrices have been reported in previous studies \cite{Isaev2010,Isaev2020}.

For RaF, and more generally for alkaline-earth-metal monofluorides, the argument for the diagonal structure of the FCFs is that fluorine, with its large electron affinity, attracts one of the $ns^2$ electrons of the AEM into its $2p^5$ orbital. The remaining valence $s$ electron of the AEM remains strongly localized around the metal atom and, therefore, does not significantly change the internuclear distance upon excitation. In contrast, in the case of R–X systems such as RaO, the bonding is bivalent; that is, the $2s^22p^4$ electronic configuration of oxygen attracts two electrons from the $7s^2$ orbital of Ra, as discussed above. The first electron is mostly localized around the oxygen atom, whereas the second is shared between the two atoms. As a result, RX molecules exhibit a polar covalent bond. The nature of the bond directly affects the prospects of these molecules for laser cooling, since the bond length changes upon radiative excitation of this electron. Further discussion of these differences is presented in the next section and illustrated in Fig.~\ref{fig:orbitals2}.

\begin{table}[h!]
\centering
\caption{Spectroscopic Constants of RaO, RaS and RaSe extracting from the MCSCF/MRCI+Q + ECP calculations. In order to compare, some values were computed using the CCSD(T)+ECP(*) and CCSD-EOM+ECP(**) method with the same basis sets and core-pseudopotentials as for the original methodology }
\begin{tabular}{@{}lccccc@{}}
\toprule
\textbf{State} & $R_e$ (\AA) & $\omega_e$ (cm$^{-1}$) & $B_e$ (cm$^{-1}$)  & $T_e$ (cm$^{-1}$) \\
\midrule
\multicolumn{6}{c}{\textbf{RaO}} \\
X$^1\Sigma^+$ & 2.101 & 593.304 & 0.2538 & 0 \\
              & 2.121* &  &    0.2490*     &  \\
              & 2.04[28] & 595.00[28] &         &  \\
a$^3\Pi$      & 2.434 & 395.227 & 0.1890 & 9714.917 \\
              & 2.468* &  &  0.1699*  &10464.130*   \\
A$^1\Pi$      & 2.465 & 439.716 & 0.1844  & 9866.91 \\
              &  &  &    &   11981.92 **\\

B$^1\Sigma^+$ & 2.315 & 532.477 & 0.2091  & 10234.01
&  \\
b$^3\Sigma^+$ & 2.422 & 390.501 & 0.1910
& 21795.36 \\
\midrule
\multicolumn{6}{c}{\textbf{RaS}} \\
X$^1\Sigma^+$ & 2.741 & 348.38 & 0.0793 & 0 \\
              & 2.770* &  &    0.0783*    &  \\
a$^3\Pi$      & 3.008 & 222.84 & 0.0659&  7846.75 \\
              & 3.032* &  &    0.0646$^{*}$  &   8000.043$^{*}$ \\
A$^1\Pi$      & 3.064 & 217.09 & 0.0635 &  7024.96 \\
              &  & &  &  9439.28 ** \\
B$^1\Sigma^+$ & 2.880 & 278.27& 0.0718  & 12753.58
&  \\
b$^3\Sigma^+$ & 2.975 & 234.73 & 0.0673
& 20839.63 \\
\midrule
\multicolumn{6}{c}{\textbf{RaSe}} \\
X$^1\Sigma^+$ & 2.894 & 151.69 & 0.0342  & 0 \\
              & 2.889$^{*}$  & &  0.0343$^{*}$   & \\
a$^3\Pi$      & 3.164 & 144.47 & 0.0286 &  5979.822 \\
              & 3.188$^{*}$  & &  0.0265$^{*}$  & 6842.09$^{*}$  \\
A$^1\Pi$      & 3.160 & 127.683& 0.0286 & 6764.435\\
              &  & & & 7422.2105 **\\
B$^1\Sigma^+$ & 3.017 & 142.912 & 0.0314  & 24039.00 &  \\
b$^3\Sigma^+$ & 3.125 & 153.11 & 0.0292 & 21829.32\\
\bottomrule
\label{tab:spectroscopy}
\end{tabular}
\end{table}

\renewcommand{\arraystretch}{1.2} 

\renewcommand{\arraystretch}{1.2} 

\subsubsection{RaO$^\pm$}

\begin{figure}[h]
\centering
\includegraphics[width=\linewidth]{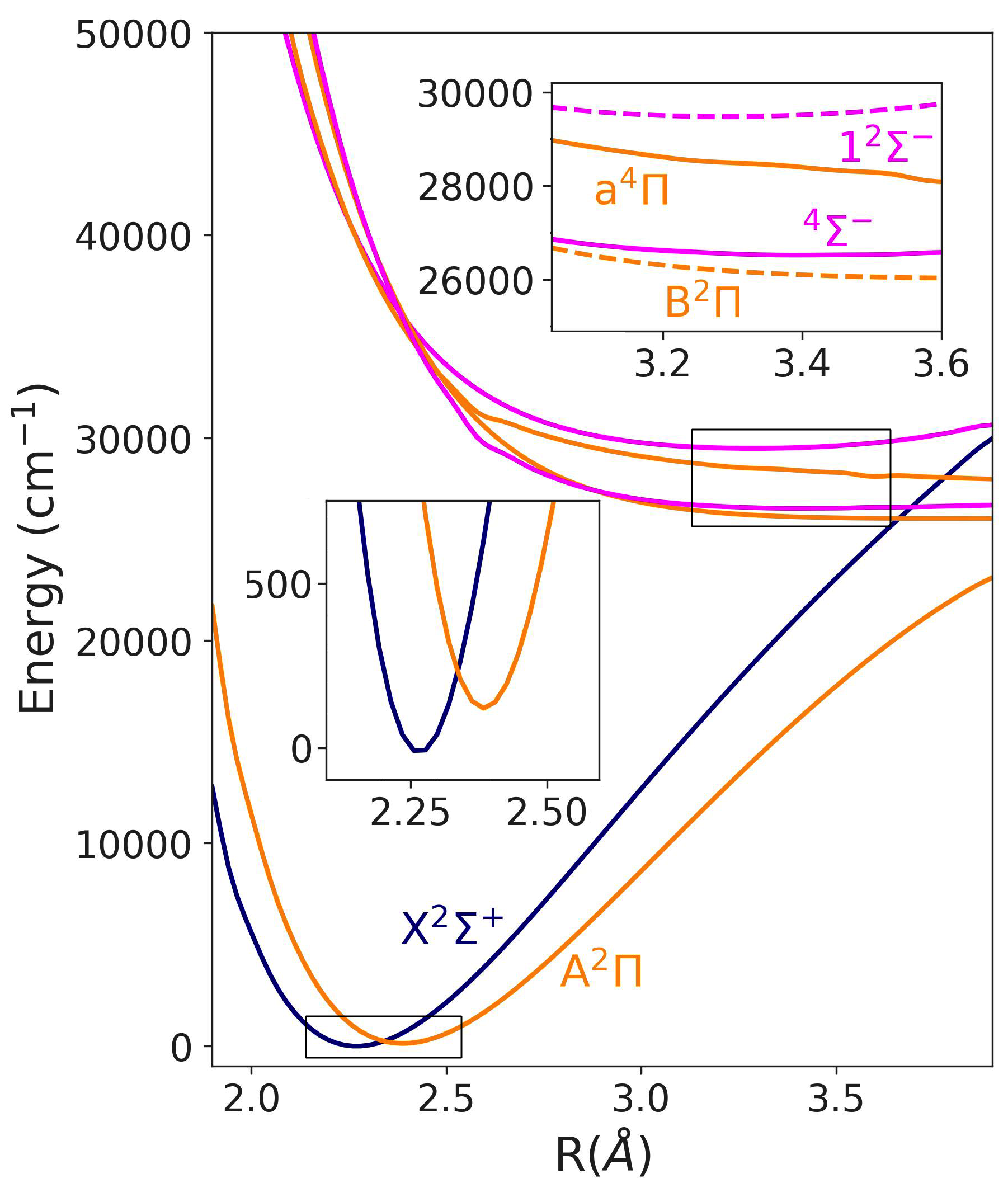}  
\caption{Potential energy curves of the lower 6 $\Lambda$-S electronic states for the RaO$^+$ molecular ion. The inset zoomed the black box regions. The axis labels in the inset are omitted because of the space, but they are the same as the big axis }
\label{fig:PES_RaO+}
\end{figure}

Figure \ref{fig:PES_RaO+} shows the lowest 6 $\Lambda -S$ electronic states of RaO$^+$. The ground state shows X$^{2}\Sigma^+$ symmetry and is only 134.245 cm$^{-1}$ below the first excited state, A$^2\Pi$. The X$^{2}\Sigma^+$ ground state has the valence electronic configuration 1$\sigma^2$1$\pi^{4}$2$\sigma^\alpha$, where the molecular orbitals are mainly described by the $2s,2p_{z}$ and $2p_{x,y}$  atomic orbitals of oxygen. The bonding is primary ionic with a strong covalent component, with the O taking the $7s$ and one $6p$ electron of Ra$^{+}$ as expected. The A$^2\Pi$ states has the electronic configuration 1$\sigma^2$1$\pi^{3}$2$\sigma^2$, so the excitation dominating the X$^{2}\Sigma^+ \longrightarrow $ A$^2\Pi$ transition is 2$\pi_x \longrightarrow$ 2$\sigma$. The $1\pi_{x,y}$ molecular orbitals are below the $2\sigma$ orbitals, so this excitation involves an internal electron instead of the most external electron, producing a notable change in the bond length when moving from the ground to the first electronic excited state, and a small energy difference.

\begin{table}[h!]
\centering
\caption{Spectroscopic Constants of RaO$^{\pm}$ extracting from the MCSCF/MRCI+Q + ECP calculations. In order to compare, some values were computed using the CCSD(T)+ECP(*) method with the same basis sets and core-pseudopotentials as for the original methodology }
\begin{tabular}{@{}lccccc@{}}
\toprule
\textbf{State} & $R_e$ (\AA) & $\omega_e$ (cm$^{-1}$) & $B_e$ (cm$^{-1}$)  & $T_e$ (cm$^{-1}$) \\
\midrule
\multicolumn{6}{c}{\textbf{RaO$^{+}$}} \\
X$^2\Sigma^+$ & 2.270 & 491.299 & 0.2173 & 0 \\
              & 2.143* &  & 0.2439*     &  \\
A$^2\Pi$ & 2.390 & 449.595  & 0.1963 & 134.245 &  \\
\midrule
\multicolumn{6}{c}{\textbf{RaO}$^{-}$} \\
X$^2\Sigma^+$ & 2.324 & 383.722 & 0.2075  & 0 \\
              & 2.285* &  &   0.2169*    &  \\
A$^2\Pi$      & 2.481  & 335.066 &0.1820 &  216.907 \\
\bottomrule
\label{tab:spectroscopy_ions}
\end{tabular}
\end{table}

\begin{figure}[h]
\centering
\includegraphics[width=\linewidth]{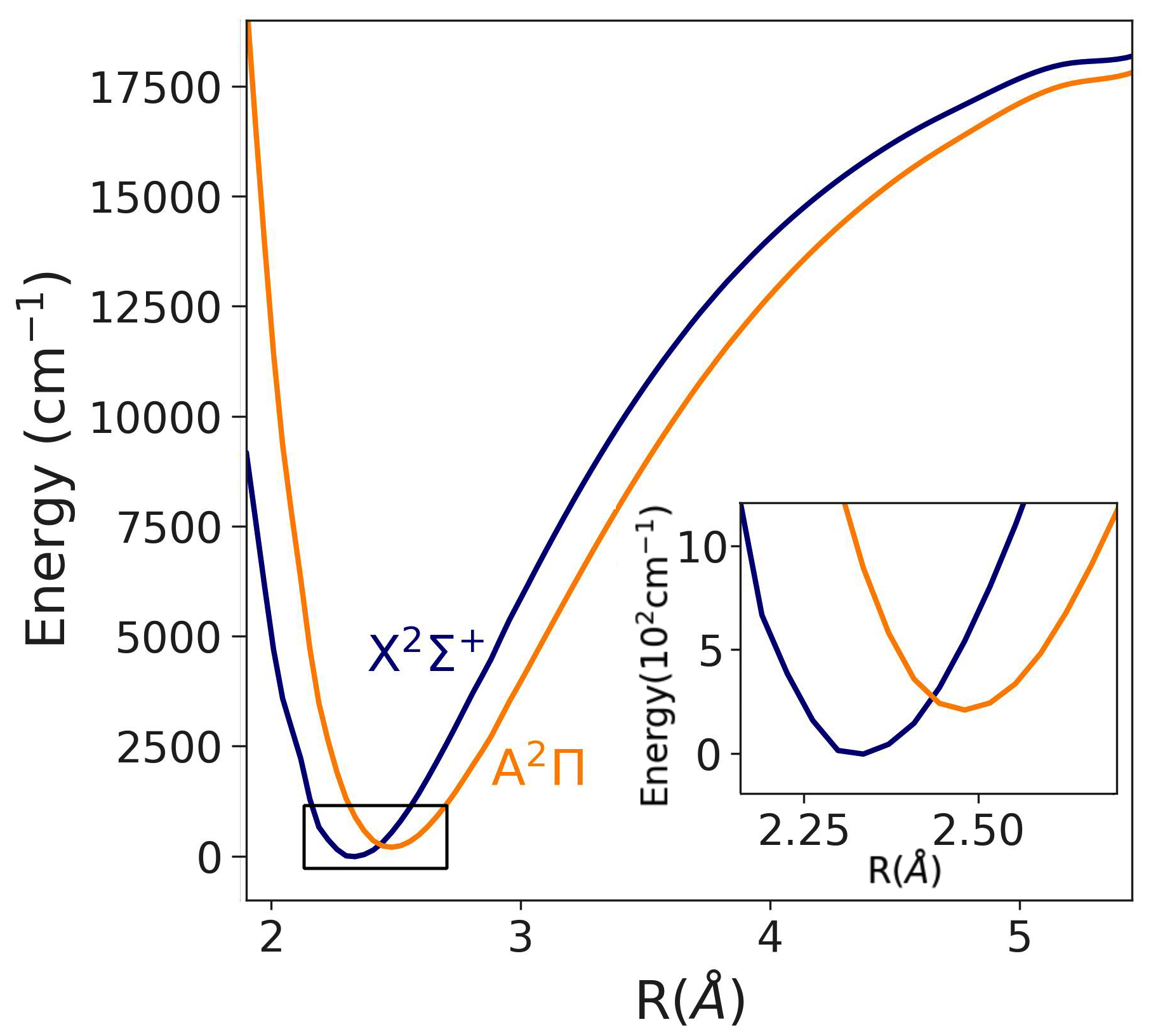}  
\caption{$\Lambda$-S potential energy curves of the ground state and first excited state of the RaO- molecular anion. The inset zoomed the black box region around the equilibrium point.}
\label{fig:RaO-_cm}
\end{figure}

For the RaO$^-$ the two electronic states of the Ra($^1$S)+O$^{-}$($^2$P$^{\circ}$) atomic channel are display in figure \ref{fig:RaO-_cm}. The electronic configuration of the incoming species in this case is similar to that of RaF, with Ra(...6$s^2 p^6$7$s^2$) and O$^-$(...2$s^2 p^5 $). But in the equilibrium still two electrons participate in the  bonding, yielding a less deep potential ground state X$^2 \Sigma^+$, than that of the other cases, neutral RaX and RaO$^+$. Again, despite the open shell structure, we see that the singlet-singlet excitation, X$^{2}\Sigma^{+} \longrightarrow $A$^{2}\Pi$, results in a considerable change in the equilibrium bonding length (see table \ref{tab:spectroscopy_ions}).

\begin{figure}[h]
\centering
 \includegraphics[width=\linewidth]{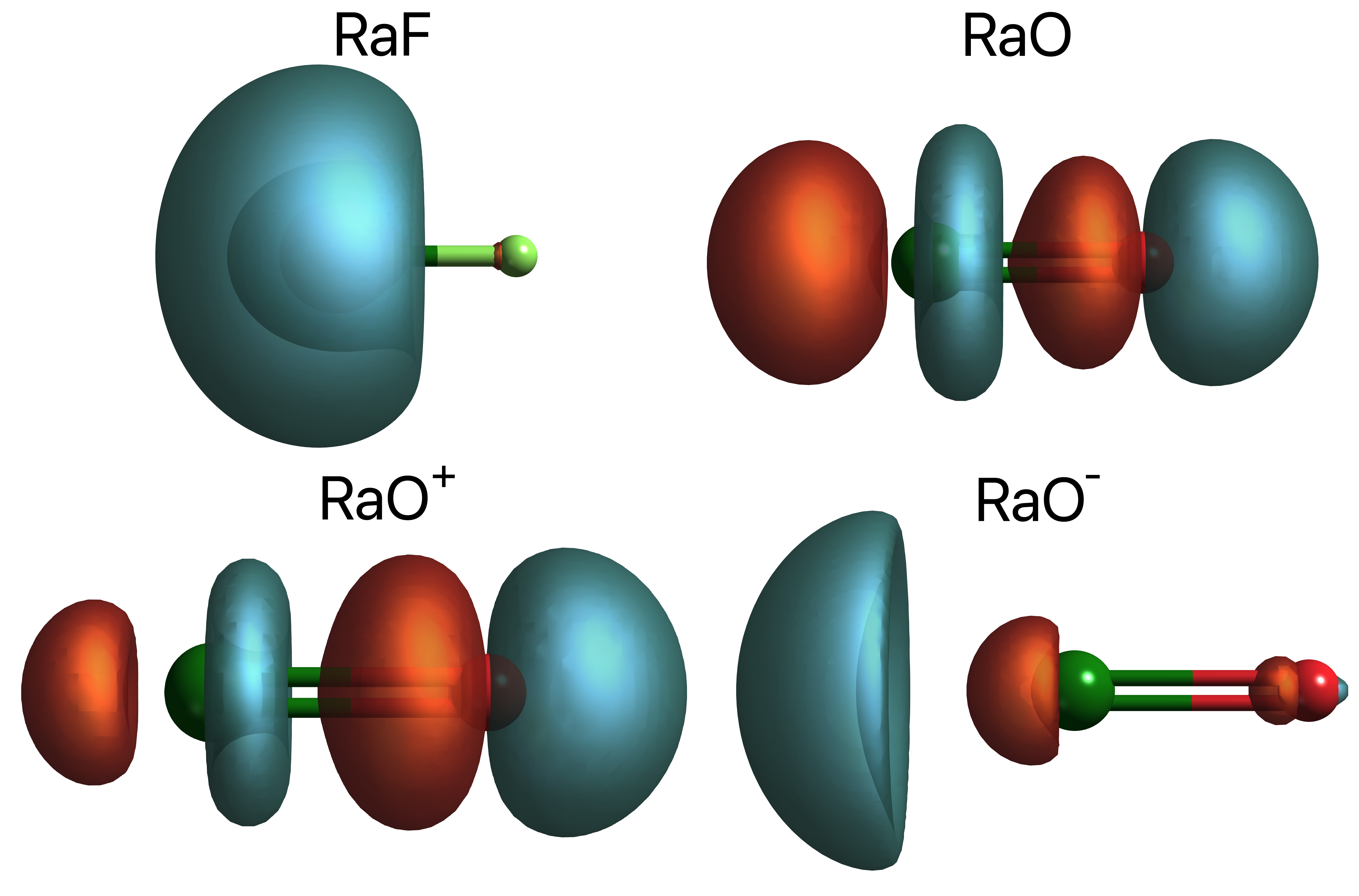}  
\caption{Single occupied molecular orbitals for the RaF, RaO$^{+}$ and RaO$^{-}$ molecules. The bonding orbital for the RaO is added for comparison purposes.}
\label{fig:orbitals2}
\end{figure}

Table~\ref{tab:spectroscopy_ions} shows the spectroscopic constants for the RaO$^{\pm}$ species. For RaO$^{+}$, we observe a difference of 0.12~\AA\ in the equilibrium bond length between the ground and first excited states. This situation is similar to that in RaX, as explained above, resulting in poor overlap of the vibrational wavefunctions for both states with the same $\nu$ quantum number, and therefore a non-diagonal Franck–Condon factor matrix. To investigate this further, we plot in Fig.~\ref{fig:orbitals2} the singly occupied molecular orbitals for the RaF and RaO$^{\pm}$ molecules. From these orbitals, the different localization patterns can be observed, which lead to significant differences in the shifts of the equilibrium bond lengths upon electronic excitation.

\subsection{spin-orbit calculations}

\begin{figure*}[htbp!]
\includegraphics[width=\textwidth]{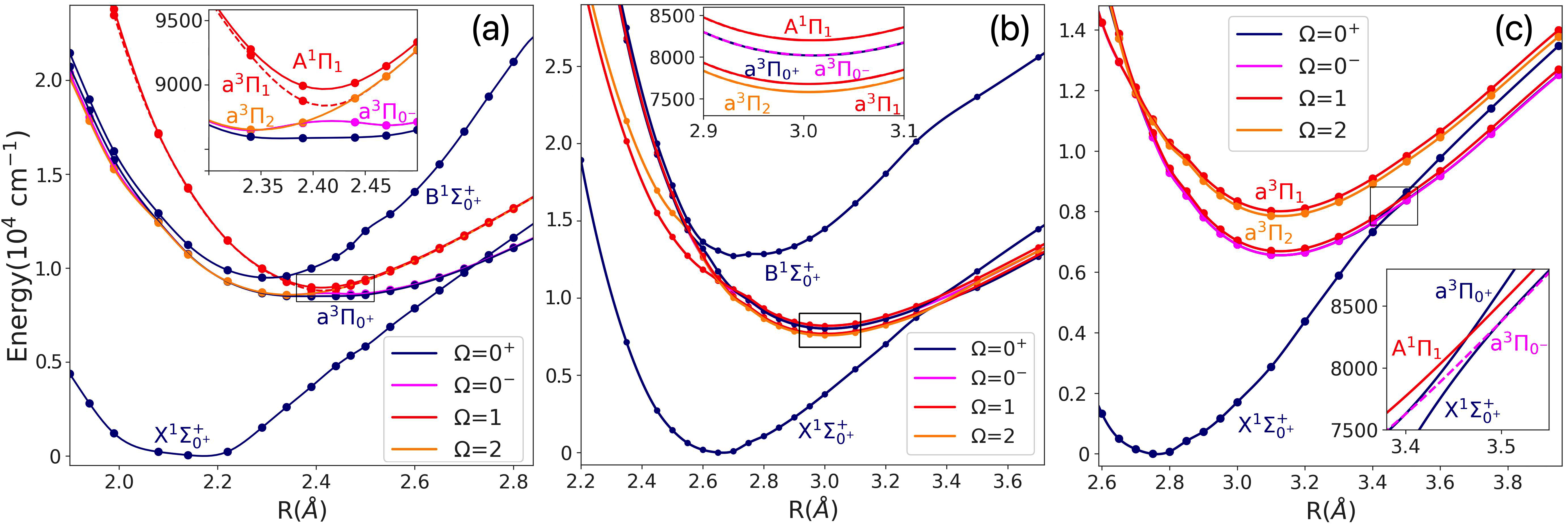}  
\caption{ The MRCI+Q + ECP + SO potential energy curves of 7  lowest $\Omega$ electronic states for the neutral Radium-monochalcogenides: (a) RaO, (b) RaS and (c) RaSe . The inset zoomed the region in the black box. The axis labels in the inset are omitted because of the space, but they are the same as the big axis}
\label{fig:PES_SO}
\end{figure*}

As mentioned before, the SO effects are included in the state interaction approach. The Pauli-Breit operator is diagonalized using the non-perturbative orbitals obtained from the MRCI+Q+ECP method. In this framework, the states are labeled by the projection $\Omega$ of the total electronic angular momentum, \textbf{J=L+S}, on the internuclear axis $\hat{R}$. In particular, the absolute value of the projection is used, $\Omega = |\Lambda+\Sigma|$. Figure \ref{fig:PES_SO} shows the resulting seven $\Omega$ states associated with the four lowest $\Lambda - S$ states for the neutral Ra-X molecules. The potential energy curves are shown for the relevant internuclear distance around the potential minima. As in the case of $\Lambda-$S states, the potential energy curves are very much alike across the different molecules.

Table~\ref{tab:spectroscopy_SO} compiles the spectroscopic constants of the $\Omega$ states. We note that relativistic effects do not significantly change these values compared to the $\Lambda - S$ states, although some contraction is observed in the $\Pi$ states, possibly due to the contraction of the $2p_{1/2}$ orbitals of oxygen. These small perturbations to the equilibrium bond length are not sufficient to improve the wavefunction overlap in the FCF maps. Regarding the energy shifts, we observe that they are on the order of hundreds of cm$^{-1}$ and, in most cases, reduce the value of $T_{e}$ for the $\Lambda - $S states, which may be a consequence of the relativistic stabilization of the $s$ and $p$ valence orbitals of the chalcogens. These values were obtained at the MRCI level, where the contribution of electronic correlation tends to be larger, as seen in Table~\ref{tab:spectroscopy} when compared with the CCSD(T)+ECP calculations.

A more detailed study of the equilibrium distances of the potential energy curves reveals that relativistic effects tend to elongate the equilibrium distances of the relevant electronic states relative to the $\Lambda -$S states. However, the contrary occurs for RaS and RaSe. In fact, the inverse behavior is observed in the vibrational harmonic frequencies of the potential energy curves. This should not come as a surprise since it has been shown that the equilibrium distance and harmonic frequencies of electronic states in diatomic molecules are correlated~\cite{Ibrahim2024}.

\begin{table}[h!]
\centering
\caption{Spectroscopic Constants of RaO, RaS and RaSe extracting from the MCSCF/MRCI+Q + ECP + SO calculations }
\begin{tabular}{@{}lccccc@{}}
\toprule
\textbf{State} & $R_e$ (\AA) & $\omega_e$ (cm$^{-1}$) & $B_e$ (cm$^{-1}$)  & $T_e$ (cm$^{-1}$) \\
\midrule
\multicolumn{6}{c}{\textbf{RaO}} \\
X$^1\Sigma^+_{0^+}$ & 2.162 & 560.125 & 0.2384 & 0 \\
a$^3\Pi_{0^{+}}$      & 2.369 & 242.255 & 0.1996 & 8501.72 \\
a$^3\Pi_{1}$       & 2.411 & 564.693 & 0.1923 & 8980.82 \\
a$^3\Pi_{2}$       & 2.346 & 415.214 & 0.2035 & 8582.193 \\
A$^1\Pi_{1}$    & 2.422 & 651.62 & 0.1925  & 8905.656  \\
B$^1\Sigma^+_{0^+}$ & 2.315 & 532.477 & 0.2091  & 10234.01
&  \\
\midrule
\multicolumn{6}{c}{\textbf{RaS}} \\
X$^1\Sigma^+_{0^+}$ & 2.691 & 438.006 & 0.0835 & 0 \\
a$^3\Pi_{0^{+}}$      & 3.010 & 244.993 & 0.0658 & 7999.123 \\
a$^3\Pi_{1}$       & 2.996 & 185.813 & 0.0664 & 7692.731 \\
a$^3\Pi_{2}$       & 3.006 & 166.848 & 0.0659 & 7650.268 \\
A$^1\Pi_{1}$      & 3.019 & 227.456 & 0.0654  & 8239.944  \\
B$^1\Sigma^+_{0^+}$ & 2.709 & 534.838 & 0.0811  & 12754.253
&  \\
\midrule
\multicolumn{6}{c}{\textbf{RaSe}} \\
X$^1\Sigma^+_{0^+}$ & 2.764 & 247.344 & 0.0374 & 0 \\
a$^3\Pi_{0^{+}}$      & 3.128& 147.089 & 0.0292 & 6567.582 \\
a$^3\Pi_{1}$       & 3.125 & 145.306 & 0.0293 & 8021.023 \\
a$^3\Pi_{2}$       & 3.126 & 144.927 & 0.0293 & 7861.956\\
A$^1\Pi_{1}$      & 3.127 & 147.156 & 0.0292  & 6700.210 \\
\bottomrule
\label{tab:spectroscopy_SO}
\end{tabular}
\end{table}


\subsection{Polarization, dipolar moments and dispersion coefficients}


\begin{table}[h!]
\centering
\caption{Calculated values for the parallel ($\alpha_{||}$) and transversal ($\alpha_{\perp}$) components of the static polarization, and permanent dipole moment $d_{0}$; for the ground state of the molecule $X^{1}\Sigma^{+}$ at the equilibrium length.} 
\begin{tabular}{@{}lcccc@{}}
\toprule
\textbf{Method} & $\alpha_{||} $ (a.u) &  \,\, $\alpha_{\perp}$ (a.u)&\,\, $d_0$ (Debye)  \\
\midrule
\multicolumn{5}{c}{\textbf{RaO}} \\
CCSD(T)+ECP & 175.604   &  48.731 & 9.1041 & \\

CCSD(T)+X2C(pVQZ-X2C)     & 185.710 &  49.075 & 8.8965 \\
CCSD(T)+X2C(dyall.av3z)     & 183.064 &  49.952 & 8.8693 \\

\multicolumn{5}{c}{\textbf{RaS}} \\
CCSD(T)+ECP & 270.893 & 93.033 &  11.3687  \\

CCSD(T)+X2C(pVQZ-X2C)    & 269.681 & 92.342& 11.2269 \\
CCSD(T)+X2C(dyall.av3z)    & 267.971 &  91.979& 11.1056\\

\multicolumn{5}{c}{\textbf{RaSe}} \\
CCSD(T)+ECP & 285.428 & 105.330 & 11.5972   \\

CCSD(T)+X2C(pVQZ-X2C)  & 290.536   &  103.719 & 11.5375  \\
CCSD(T)+X2C(dyall.av3z)     & 287.311 &  105.072 & 11.6134 \\

\bottomrule
\label{tab:dipolar}
\end{tabular}
\end{table}

Table~\ref{tab:dipolar} reports the parallel and perpendicular components of the static dipole polarizability, together with the permanent dipole moments, obtained using the different computational methodologies considered in this work. For the CCSD(T)+X2C calculations performed with the \textsc{DIRAC} code, only results obtained with the dyall.av3z basis set are shown, although calculations with dyall.av2z were also carried out and yielded consistent results. Overall, a very good level of agreement is observed among the three computational approaches. The differences between the two X2C-based calculations arise not only from the use of different basis sets, but also from the fact that the pVQZ-X2C calculations explicitly include full core-valence correlation effects. In contrast, these are not fully accounted for in the dyall.av3z calculations.

With respect to the permanent dipole moments, we emphasize the remarkably large values obtained for these systems, in particular for RaS and RaSe, which exceed $10~\mathrm{D}$. In a previous study based on a machine-learning predictor, the largest dipole moment reported for a diatomic molecule was predicted to be approximately $11.52~\mathrm{D}$. For the same molecule, high-level \textit{ab initio} calculations using the CCSD(T) method with small-core pseudopotentials and consistent Dunning-type basis sets yielded a value of $11.73~\mathrm{D}$, which is comparable to the dipole moments obtained here for RaS and RaSe. Although the dipole moment of RaO is smaller than those of RaS and RaSe, it remains exceptionally large when compared with other diatomic molecules, such as heteronuclear alkali-metal dimers (typically $< 5~\mathrm{D}$)~\cite{Ladjimi2024}, or radium monohalides such as RaF and RaCl~\cite{Ahmed2025}.

The large permanent dipole moment of RaO (exceeding $2~\mathrm{D}$), and similarly of RaS and RaSe, suggests the possible existence of an excited dipole-bound state (DBS) in the RaO$^{-}$ anion~\cite{Crawford1967}. Dipole-bound states in BeO$^{-}$ have been explored previously~\cite{macaritolo2017}, where notable features such as a large overlap with the ground electronic state were reported. Similar phenomena may arise in the RaX systems, making the investigation of dipole-bound states in these molecules an interesting prospect for future work.

The remarkable difference between the parallel and perpendicular components of polarization is the result of the large contributions from the low energy transitions $\sigma\rightarrow\sigma$ and $\sigma\rightarrow\pi_{0}$ in the lowest excited transitions. In general, and in contrast with the values for the dipole moment, the mean polarizations of the RaX molecules here are smaller than those for the heteronuclear alkaline dimers. For instance, LiNa has a mean polarizability of 237.7 a.u and RbCs has 638.6 a.u. other heteronuclear alkaline dimers possess values bounded between these two \cite{Zuchowski2013}.

\begin{table}[h!]
\centering
\caption{Dispersion coefficient $C_{6}^{\rm{disp}}$ for the interaction between the study Radium-monochalcogenides. The values are presented in atomic units.}
\begin{tabular}{@{}lcccc@{}}
\toprule
\textbf{Molecule} & \,\, RaO \,\, &  \,\, RaS\,\, &\,\, RaSe \,\, \\
\midrule
RaO &\,\, 1394.853 \,\, &   &  & \\

RaS & \,\,2163.358 \,\,  & \,\, 3378.980 \,\, &  & \\

RaSe & \,\,2349.168\,\,   & \,\, 3669.705\,\, &\,\,3985.470 \,\,& \\
\bottomrule
\label{tab:C6}
\end{tabular}
\end{table}

Finally, for completeness, we have computed the dynamic polarizability at imaginary frequencies and estimated the dispersion coefficient through equation \ref{eq:eq4}. The obtained values are tabulated for the different molecules in Table \ref{tab:C6}.
Once again, comparing with the heteronuclear alkaline dimers, and as expected from the difference in the mean polarizability, the dispersion coefficients are smaller for the RaX molecules.

\section{Conclusions}

We have explored the electronic properties of the radium monochalcogenides RaX (X = O, S, Se) and the RaO$^{\pm}$ ions, including their lowest electronic states. In addition to potential energy curves, we calculate the electric dipole moments, dipole polarizabilities, and dispersion interaction coefficients of these molecules using a variety of quantum-chemical methods. The potential energy curves were obtained using the MRCI+Q+ECP+SO approach, in which electronic correlation was treated at the MRCI+Q level. Scalar relativistic effects were included through the use of small-core, energy-consistent relativistic pseudopotentials, while spin–orbit contributions were incorporated perturbatively using the state-interaction method. Electric dipole moments and dipole polarizabilities were computed using coupled-cluster theory with single, double, and perturbative triple excitations. Relativistic effects were included either through the exact two-component Hamiltonian, CCSD(T)+X2C, or via pseudopotentials, CCSD(T)+ECP. Additional calculations using the EOM-CCSD+ECP method were performed to benchmark the transition energies obtained with the MRCI+Q+ECP+SO approach.

We find that the RaX molecules, like other alkaline-earth-metal monochalcogenides, exhibit predominantly divalent bonding character that is not significantly modified by relativistic effects. The corresponding ground states, $X^{1}\Sigma^{+}$, correlate with the first excited atomic dissociation channel, and the lowest excited electronic state lies below $9000\ \mathrm{cm^{-1}}$ in all cases. Due to the divalent bonding character and the closed-shell electronic structure of these molecules, a significant shift in the equilibrium bond length occurs upon electronic excitation. This results in non-diagonal Franck–Condon factors, making these species unsuitable for laser cooling. Nevertheless, the resulting electronic structure leads to very large permanent dipole moments, reaching values of approximately 11.3 and 11.5 Debye for RaS and RaSe, respectively, which may be advantageous for manipulation with external electric fields.

For the ionic species, very small excitation energies are found between the ground and first excited electronic states. This behavior arises because the dominant contribution to the excitation corresponds to an atomic-like $np_x \rightarrow np_z$ transition, which is relatively low in energy. Despite the open-shell character of these ions, the singly occupied orbitals retain strong covalent character, in contrast to the situation in laser-coolable radium monohalides.

Our results show that pseudopotential-based methods perform well overall compared with more sophisticated relativistic approaches, both for the systems studied here and for additional benchmarking calculations reported in the Appendix. Nevertheless, further refinements based on fully four-component relativistic calculations could still improve the accuracy of the potential energy curves and transition energies.

We anticipate that the synthesis of these molecules can be achieved via laser ablation of a solid radium target in an atmosphere containing a chalcogen donor gas. For instance, one possible scenario would involve providing an inlet for injecting O$_2$. This idea is motivated by recent results on CaH formation via the reaction Ca + H$_2$ $\longrightarrow$ CaH + H~\cite{Sun2026}. Although the reaction Ra + O$_2$ $\longrightarrow$ RaO + O is exothermic, vibrational excitation of the hot Ra atoms may facilitate the formation of RaO, which could subsequently be cooled through collisions with a helium buffer gas.

\section{Acknowledgments}

The authors thank Ronald Fernando Garcia Ruiz and Silviu-Marian Udrescu for fruitful discussions and motivating some of the systems studied here. The authors knowledge the support by the AFOSR Grant No. FA9550-23-1-0202.

\section{Appendix A}
\label{sec:appendix}

In this section, we estimate the accuracy of the potential energy curves and transition energies $T_e$ of the MRCI+Q+ECP+SO approach by comparison with a fully relativistic methodology. To this end, we benchmark our approach against the well-studied RaF molecule by comparing our results with those obtained in previous studies employing more sophisticated relativistic methods.
In addition, we assess the performance of the CCSD(T)+ECP method for the calculation of dipole moments of SrF, BaF, and RaF, for which high-level theoretical data and experimental measurements are available \cite{ERNST1985,torring1986,Haldar2021}. We also evaluate the accuracy of transition energies computed using the EOM-CCSD+ECP method by comparing our results with previously reported experimental and theoretical values \cite{Revelli1975,GOTTSCHO1980,Field1974,KHATIB2018}.

\subsection{RaF PEC}

Here we employ the MRCI+Q+ECP+SO method to compute the lowest electronic states of the RaF molecule. In particular, we consider the $X^{2}\Sigma^{+}{1/2}$, $^{2}\Pi{1/2}$, and $^{2}\Pi_{3/2}$ states. Following the methodology described above, we use the ECP78MDF pseudopotential together with the aug-cc-pVQZ-PP basis set to describe the Ra atom, while the aug-cc-pwCV5Z basis set is used for F.

In the orbital space, the $6s$, $6p$, $6d$, and $7p$ orbitals of Ra are included. The $6s$ and $6p$ orbitals are treated as closed, while the remaining orbitals are included in the active space. For F, the $1s$ orbital is included as a closed orbital, whereas the $2s$, $2p$, and $3s$ orbitals are included in the active space. In the $C_{2v}$ point group, the total orbital space corresponds to (10,4,4,1), while the active space is (7,3,3,1).

\begin{figure}[h]
\centering
 \includegraphics[width=\linewidth]{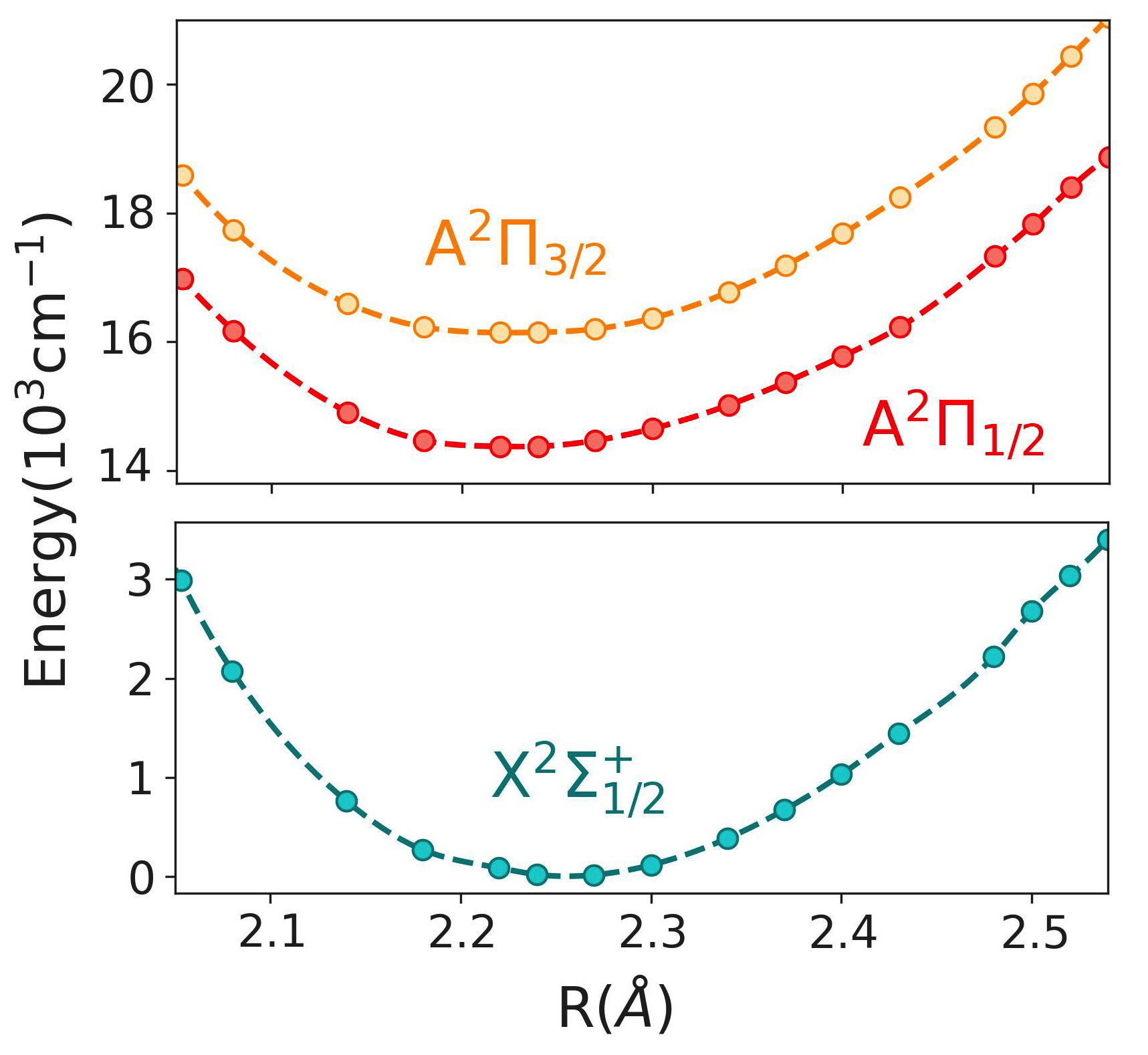}  
\caption{Lowest electronic states of the RaF molecule computed in the MRCI+Q+ECP+SO approach}
\label{fig:RaF_states}
\end{figure}

The resulting potential curves for the X$^{2}\Sigma^{+}_{1/2}$, $^{2}\Pi_{1/2}$ and $^{2}\Pi_{3/2}$ electronic states are shown in Fig.~\ref{fig:RaF_states} and the computed spectroscopy constants are shown in table \ref{tab:RaF} along with values for comparison.

\begin{table}[h!]
\centering
\caption{Spectroscopic constants of the lowest electronic states of RaF computed using different theoretical approaches, including the MRCI+Q+ECP+SO method employed in this work.}
\begin{tabular}{@{}lcccccc@{}}
\toprule
\textbf{State} & $R_e$ (\AA) & $\omega_e$ (cm$^{-1}$) & $B_e$ (cm$^{-1}$)  & $T_e$ (cm$^{-1}$)& Ref. \\
\midrule
X$^2\Sigma^+_{1/2}$ & 2.25 & 490.125 & 0.2384 & 0 & This Work \\
 & 2.244 & 428 &  &  & \cite{Isaev2010} \\
$^2\Pi_{1/2}$      & 2.24 & 442.255 & 0.1996 & 14410 & This work \\
     & 2.248 & 436.8 &   & 13320 & \cite{Zaitsevskii2022} \\
     & 2.244 & 432 &   & 14012& \cite{Isaev2010}\\
     & & 435.5 &   & 13288 & Exptl. \cite{Ruiz2019}\\

$^2\Pi_{3/2}$       & 2.23 & 464.693 & 0.1923 & 16142.630 & This work \\
      & 2.243 & 436.3 &  & 16644 & \cite{Zaitsevskii2022} \\
\bottomrule
\label{tab:RaF}
\end{tabular}
\end{table}

Equilibrium bond lengths are generally in good agreement across all calculations, while the vibrational frequencies are, on average, overestimated by about 30 cm$^{-1}$, corresponding to less than 2\%. The excitation energies are more sensitive to the computational methodology. For the $X^{2}\Sigma^{+}{1/2} \rightarrow {}^{2}\Pi{1/2}$ transition, the deviation from the experimental value is 8.44\%, which is smaller than that obtained in previous theoretical calculations. For the second transition, $X^{2}\Sigma^{+}{1/2} \rightarrow {}^{2}\Pi{3/2}$, the error is significantly smaller, amounting to 3.72\%, and is also reduced compared to earlier theoretical results.

The errors with respect to the experimental values are below 9\%, which gives us confidence in the reported results. We note that these transition energies are obtained from the MRCI+Q treatment of the electronic wavefunction; however, we also report EOM-CCSD+ECP values, which may be closer to experiment due to their improved treatment of dynamic electron correlation.

\subsection{Dipole moment of SrF,BaF and RaF}

We implemented the finite-field difference method, with the energies computed at the CCSD(T)+ECP level, to obtain the dipole moment of the heavy AEM monofluorides (Sr-Ra)F.

For the AEM atoms we use the small-core fully-relativistic pseudopotentials including 28,46 and 78 core electrons for Sr,Ba and Ra respectively, along with the aug-cc-pVQZ-PP basis set. For the F atom, we use the aug-cc-pwCV5Z basis set. We also included full core-valence correlations for those electrons not included in the pseudopotential.

\begin{table}[h!]
\centering
\caption{Dipole moments of SrF, BaF, and RaF computed using the finite-field method with the 10-electron ECPXMDF pseudopotentials at the CCSD(T) level of theory. Additional experimental values and results from fully relativistic calculations are included for comparison.}
\begin{tabular}{@{  }lccc@{}}
\toprule
 \hspace{10 pt}\textbf{Molecule} & \hspace{10 pt} Dipole moment (D) & Ref. \\
\midrule
\midrule
 \hspace{20 pt}SrF & 3.501 & This Work \\
    & 3.468 &   Exptl. \cite{ERNST1985}\\
 \hspace{16 pt} BaF & 3.220 & This Work \\
    & 3.170 & Exptl. \cite{torring1986}\\
 \hspace{16 pt} RaF & 3.923 & This work \\
    & 3.85 & \cite{Haldar2021} \\
\bottomrule
\label{tab:dipoles_test}
\end{tabular}
\end{table}

Table \ref{tab:dipoles_test} shows the computed dipole moments together with experimental and theoretical reference values. The results obtained for SrF and BaF are in good agreement with the experimental measurements, with relative errors not exceeding 1$\%$. For RaF, no experimental value is currently available. However, Ref.~\cite{Haldar2021} reports a value based on relativistic coupled-cluster calculations. The relative difference between that value and our result is only 1.9$\%$. This overall good agreement between the CCSD(T)+ECP dipole moments and the available reference data gives us confidence in the predicted values for the radium monochalcogenides.

\subsection{Transition energies for BaO}
Finally, we test the performance of the EOM-CCSD+ECP method for computing the transition energies into the lower electronic states of the BaO system, for which both experimental data and theoretical calculations based on quasi-relativistic pseudopotentials are available.

\begin{table}[h!]
\centering
\caption{Equilibrium distance and transition energies of the BaO two lower electronic states. Experimental and other theoretical values are presented for comparison of the current values }
\begin{tabular}{@{  }lcccc@{}}
\toprule
 \textbf{States} & \hspace{10 pt} R$_{\rm{eq}} ($\AA$)$ & T$_{e}$ (cm$^{-1}$) & Ref.  \\
\midrule
\midrule
A$^{1}\Pi$ & 2.292 & 18136& This work \\
&2.361 &14963  & \cite{KHATIB2018}\\
& 2.292  & 17619.7 & Exptl. \cite{GOTTSCHO1980} \\
& 2.290  & 17568 & Exptl. \cite{Field1974} \\
B$^{1}\Sigma^{+}$ & 2.134 & 17229& This work \\
&2.268 &15572  & \cite{KHATIB2018}\\
& 2.1335  & 16807.6 & Exptl. \cite{GOTTSCHO1980} \\
& 2.134  & 16722 & Exptl. \cite{Field1974} \\

\bottomrule
\label{tab:BaO_test}
\end{tabular}
\end{table}

We use the ECP46MDF/aug-cc-pVQZ-PP set for modeling the Ba electrons, and for the Oxygen the same aug-cc-pwCV5Z, to describe $s,p,d,f$ and $g$ functions of the eight electrons. Full core-valence correlations are included between the 10 free electrons of Ba and the 8 electrons of oxygen. Table \ref{tab:BaO_test} summarizes the results. Two main conclusions can be drawn from these values. First, the EOM-CCSD+ECP method, using the ECP46MDF pseudopotential for Ba, provides excellent agreement with experimental transition energies. Nevertheless, a slight systematic overestimation of the transition energies is observed, even for the cases considered here, which correspond predominantly to single-excitation transitions. This tendency has also been observed in calculations on other molecules studied in our group.

The second aspect concerns the comparison with other theoretical values reported in the literature. These were computed at the MRCI level using the non-relativistic ECP46MHF pseudopotential together with the custom ECP46MHF$\_$MP2 basis set. In that work, the oxygen atom was described using the highly accurate aug-cc-pV6Z basis set. Therefore, the differences between our results and those reported in Ref. \cite{KHATIB2018} mainly arise from the use of the fully relativistic ECP46MDF pseudopotential in the present calculations, leading to energy differences of more than 2000 cm$^{-1}$ due to the right treatment of inner-electron relativistic effects of Ba.
\newpage
\bibliographystyle{apsrev4-2} 
\bibliography{main}
\end{document}